\documentclass[pre,aps,twocolumn]{revtex4}
\usepackage{graphicx}
\usepackage{natbib}
\usepackage{color}
\usepackage{amsmath}
\def\strutdepth{\dp\strutbox}
\def\nw#1{\strut\vadjust{\kern-\strutdepth\vtop to0pt{\vss\hbox to\hsize
{\hskip\hsize\hskip5pt$\leftarrow$\hss\strut}}}{\em #1}}

\newcommand{\dx}{\textrm{d}x}
\newcommand{\dy}{\textrm{d}y}
\newcommand{\dz}{\textrm{d}z}
\newcommand{\sref}{^\textrm{ref}}
\newcommand{\sel}{^\textrm{el}}

\begin{document}

\title{Capillarity of soft amorphous solids: a microscopic model for surface stress}

\author{Joost H. Weijs$^1$, Jacco H. Snoeijer$^1$, and Bruno Andreotti$^2$}
\affiliation{
$^{1}$Physics of Fluids Group and MESA$^+$ Institute for Nanotechnology, 
University of Twente, P.O. Box 217, 7500 AE Enschede, The Netherlands\\
$^{2}$Physique et M\'ecanique des Milieux H\'et\'erog\`enes, UMR
7636 ESPCI -CNRS, Univ. Paris-Diderot, 10 rue Vauquelin, 75005, Paris
}

\date{\today}%

\begin{abstract}
The elastic deformation of a soft solid induced by capillary forces crucially relies on the excess stress inside the solid-liquid interface. While for a liquid-liquid interface this ``surface stress" is strictly identical to the ``surface free energy", the thermodynamic Shuttleworth equation implies that this is no longer the case when one of the phases is elastic. Here we develop a microscopic model that incorporates enthalpic interactions and entropic elasticity, based on which we explicitly compute  the surface stress and surface free energy. It is found that the compressibility of the interfacial region, through the Poisson ratio near the interface, determines the difference between surface stress and surface energy. We highlight the consequence of this finding by comparing with recent experiments and simulations on partially wetted soft substrates. 
\end{abstract}

\maketitle

\section{Introduction}
In the last decades, surface effects in solid state physics have been widely investigated from the applied and fundamental point of view. An appropriate thermodynamic framework has been developed to describe the interplay between elasticity and surface effects in crystalline solids \cite{BGIE97,FF97,WK77,I04,MS04,AVMJ88,MSM05,R02}. An essential result is that the concept of surface tension cannot be applied without any caution to a solid condensed phase. This description is usually, however, not considered in the soft matter community, which has focused on problems of adhesion, elasto-capillarity of slender bodies, deformation of an elastomer by a liquid drop, etc. \cite{BicoNATURE,PRDBRB07,BoudPRE,PyEPJST,HonsAPL,RomanJPCM,ChiodiEPL,HurePRL,L61,Rusanov75,Yuk86,Shanahan87,CGS96,White03,PericetCamara08,PBBB08,MPFPP10,SARLBB10,LM11,JXWD11,DMAS11,SBCWWD13,WBZ09,Style12,Limat12}.

\begin{figure}[ht!]
\includegraphics{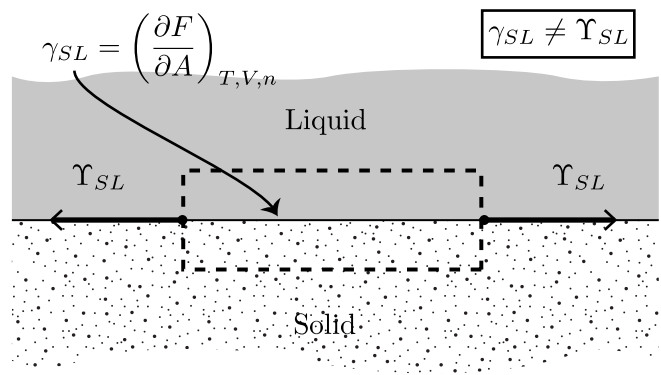}
\caption{\label{fig:intro} Schematic of a planar interface between a liquid and solid phase.
The surface free energy associated with this interface is $\gamma_{SL}=(\partial F/\partial A)_{T,V,n}$. 
In the case of a solid-liquid interface this surface free energy is generally {\em not} equal to the surface	 stress (force per unit length) at the interface $(\Upsilon_{SL})$ indicated by the arrows at the dashed control volume.}
\end{figure}

As already pointed out in \cite{WAS13} a correct treatment of elastocapillarity requires to distinguish between two excess interfacial quantities: the surface energy and the surface stress, cf.~Fig.~\ref{fig:intro}. 
A thermodynamical approach to derive the relation between surface energy and surface stress yields the well-known Shuttleworth-relation \cite{Shuttleworth50}:
\begin{equation}
\label{POEPIE}
\Upsilon_{\rm AB}=\frac{\textrm{d}\gamma}{\textrm{d}\varepsilon}+\gamma\;.
\end{equation}
with $\varepsilon$ the elastic strain parallel to the interface.
From this relation it is immediately clear that for incompressible liquids, and {\em only} for such systems, the surface energy and surface stress are equal and are usually called ``surface tension''. 
The aim of this paper is to adapt this framework previously proposed for crystalline solids, to the case of elastomers and gels: We will extend the concept of surface stress to the case of two soft condensed phases in contact. 
Specifically, we will propose a microscopic model for these macroscopic excess quantities, in the case of soft elastic amorphous solids, based on the density functional theory in the sharp interface approximation~\cite{MK92,GD98,BauerEPJ,WMALS11}. 
We will show that the compressibility of the superficial layer, quantified by the Poisson ratio $\nu$, is the key characteristic of elastomers for elasto-capillary effects. The central result will be that the solid-liquid surface stress $\Upsilon_{SL}$ relates to the interfacial energies in the following way:
\begin{equation}
\label{eq:upsilon_SL}
\Upsilon_{SL} = \frac{\nu}{1-\nu} \gamma_{SL} + \frac{1-2\nu}{1-\nu}\left( \gamma_{SV} + \gamma_{LV} \right).
\end{equation}
If the surface layer is perfectly incompressible, i.e. $\nu=1/2$, this gives a strict equality between surface energy and surface stress: $\Upsilon_{SL}=\gamma_{SL}$, as for liquid interfaces. The other extreme limit is that the stress directions perfectly decouple, i.e. $\nu=0$, for which $\Upsilon_{SL}=\gamma_{SV}+\gamma_{LV}$. 

In Sect.~\ref{sect:micmac} we derive a microscopic model to describe fluid and solid behaviour, which we use to derive $\Upsilon_{SL}$ in Sect.~\ref{sect:forces}.
The consequence of this result [Eq.~\ref{eq:upsilon_SL}] will be discussed in detail in the discussion of Sect.~\ref{sect:disc}, where we compare our findings to recent experiments and numerical results.

\section{Microscopic model}
\label{sect:micmac}
\subsection{Microscopic and macroscopic stress in a condensed phase}

We wish to define a model to compute the mechanical stresses that arise in the vicinity of a fluid-fluid or fluid-solid interface. These surface or capillary forces originate from molecular interactions, and predicting their strength thus calls for a description at that scale. Before developing this in detail, let us first consider the purely macroscopic viewpoint and define a macroscopic (or thermodynamic) stress tensor $\bar\Sigma$. In the absence of bulk forces, the mechanical equilibrium condition is
\begin{equation}
\nabla \cdot \bar\Sigma = 0.
\end{equation}
For a planar interface whose normal points along the $z$-direction, this implies that the normal stress $\Sigma_{zz}$ is constant and is identical in both phases (assuming homogeneity in the $x,y$-directions). When the interface is curved,  the boundary condition at the interface involves a normal stress discontinuity, to account macroscopically for the microscopic interactions in the surface layers. For a fluid-fluid interface, this can be expressed as the difference in thermodynamic pressure, $\Delta P = \gamma \kappa$, where $\gamma$ is the surface tension and $\kappa$ is the curvature~\cite{bookDeGennes}.

This macroscopic result can be contrasted with a purely microscopic approach that explicitly includes the molecular interactions. Microscopically, the interfacial region will be continuous, but one should still recover \emph{as a result} that the bulk stress far away from a curved interface is different in the two phases. For fluid-fluid interfaces, such a microscopic framework has been developed based on Density Functional Theory using the so-called sharp-kink approximation~\cite{MK92,GD98}. As will be further outlined in the next subsection, the key assumption of this theory is that the molecular interactions can be split in a long-range attractive part and a short-ranged repulsive interaction~\cite{WMALS11}. The attraction is described by a (mean-field) potential $\phi$, while the repulsion is a contact interaction described by a microscopic stress tensor $\bar\sigma$. The mechanical equilibrium equation then reads
\begin{equation}
\label{eq:miceq}
\nabla \cdot \bar\sigma - \nabla \phi= 0.
\end{equation}
The sharp-kink approximation assumes that the interface between the two phases is perfectly sharp, but that the interaction potential is long-ranged and extends across the interface -- most notably, the potential varies only near the interface and $\nabla \phi$ directly accounts for interfacial stresses. Since capillary effects are already included in this description, the boundary condition is now the \emph{continuity} of the normal stress $\sigma_{zz}$ across the interface. This model is very close to the original work by Laplace \cite{rowlinson}, who postulated an internal pressure to balance the strong attractive forces between molecules. For incompressible liquids near a straight interface, we will derive the following connection between macroscopic and microscopic stress:
\begin{equation}
\label{eq:micmac}
\Sigma_{ij} = \sigma_{ij} - (\phi+\phi_0) \delta_{ij}\;.
\end{equation}
where $\phi_0$ is a constant. One should bear in mind that even under atmospheric conditions, the repulsive stress $\sigma_{ij}$ is orders of magnitude larger than $\Sigma_{ij}$, as it balances the strong attractive interaction $\phi$.
An estimate for the repulsive pressure is obtained by $\gamma/a$, where $a$ is the scale of the repulsion (a few Angstroms) and thus $p_r=-\sigma_{ii} \sim 10^{8}$ N/m$^2$, as already estimated by Laplace \cite{rowlinson}. 

The description above is valid for fluids. Below, we will generalize this formulation to account for elastic solids whose elastic properties arise from entropic effects. In other words, the description focuses on solids such as gels and elastomers, which are polymer solutions in which there is no connected structure based on enthalpic (chemical) elastic effects. The goal is to derive an expression for the surface stress [Eq.~\eqref{eq:upsilon_SL}] from the microscopic interactions.

\subsection{Density functional theory in the sharp interface approximation for liquids}
We consider a simple model of liquid based on a pair potential $\varphi$. The key idea of the Density Functional Theory is to express the grand potential $\Omega=U-TS-\mu N=F-\mu N$ as a functional of the particle density $\rho$ and to perform a functional minimisation for given values of the chemical potential $\mu$ and of the temperature $T$. The free energy $F$ is not known explicitly and needs to be built by integration, starting from the free energy of a known state. For a perturbation theory, one may choose hard spheres or the same system with only a repulsive potential, and then treat the van der Waals attraction as a corrective effect. 

In the sharp interface approximation, one uses a standard van der Waals expansion of the free energy around the hard-sphere reference system (whose pair potential is denoted $\varphi_{HS}(r)$), computed in the local density approximation.
\begin{eqnarray}
F[\rho]&=&\int f_{HS}(\rho(\vec r)) d\vec r\\
&+&\frac{1}{2} \int_0^1 d\lambda \int d\vec r_1\int d\vec r_2 \rho(\vec r_1)\rho(\vec r_2) g_\lambda(\vec r_1,\vec r_2)\varphi(|\vec r_2-\vec r_1|)\nonumber
\end{eqnarray}
where $g_\lambda$  is the pair correlation function in a system of same geometry and same volume, for which the interaction is $\varphi_\lambda(r)=\varphi_{HS}(r)+\lambda [\varphi(r)-\varphi_{HS}(r)]$ above the sphere radius. One can use, for instance, a low density approximation for $g_\lambda$:
\begin{equation}
g_\lambda(r)\sim \exp(-\varphi_\lambda(r)/k_BT)
\end{equation}
where $k_B$ is Boltzmann's constant and $T$ the temperature. We now apply the Gibbs interface idealization and assume that the density is homogeneous on both sides of the interface. We furthermore extend the description to several phases and introduce $\phi_{\alpha\beta}(\vec{r})$ the effective potential associated to the influence of phase $\alpha$ on phase $\beta$, at $\vec{r}$:
\begin{equation}\label{eq:phi}
\phi_{\alpha\beta}(\vec{r}) = \rho_\alpha \rho_\beta \int_{\cal V_\alpha} d\vec{r}^\prime \varphi_{\alpha\beta}(|\vec{r}-\vec{r}^\prime|) g_{r,\alpha\beta}(|\vec{r}-\vec{r}^\prime|)\;.
\end{equation}
Note that in general $\phi_{\alpha \beta}\neq \phi_{\beta \alpha}$, unless the domains of $\alpha$ and $\beta$ have an identical shape. Altogether, we get contribution to the free energy associated to the phase $\alpha$ as:
\begin{equation}
F_\alpha=\int_{\cal V_\alpha} \left[p_r +\frac{1}{2} \phi_{LL}(\vec r)+\phi_{SL}(\vec r)\right] d\vec r
\end{equation}
The long-range attractive interaction between liquid and solid molecules gives rise to a potential $\phi_{SL}$ i.e. an energy per unit volume. The short-range repulsive interactions are assumed to lead to an isotropic stress i.e. to  a contact pressure $p_r$~\cite{HansenB}.

Minimising the free energy, we obtain the (microscopic) equilibrium condition inside the liquid,
\begin{equation}
\label{eq:liquid}
\nabla \left( p_r+\phi_{LL}+\phi_{SL} \right) = 0
\end{equation}
so that the total potential $p_r+\phi_{LL}+\phi_{SL}$ must be homogeneous \cite{GD98, MK92}.  This is the same result as obtained in Eq.~\eqref{eq:micmac}, as for liquids $\bar\sigma=-p_r$ and the macroscopic stress $\bar\Sigma$ is homogeneous in the liquid phase.

In this work, we work with planar interfaces only and therefore, for future reference, we define the potential $\phi$ in phase $\alpha$ due to the presence of a semi-infinite phase $\beta$ at distance $h$ from an interface as:
\begin{equation}
\Pi_{\alpha\beta}(h)=\int_{-\infty}^\infty \int_{-\infty}^\infty \int_{h}^\infty \phi_{\alpha\beta}(|\vec{r}|) \dz \dy \dx \;,
\end{equation}
Note that, altough in general $\phi_{\alpha\beta}\ne\phi_{\beta\alpha}$ the equivalence $\Pi_{\alpha\beta}=\Pi_{\beta\alpha}$ does hold, since in that case the phases $\alpha$ and $\beta$ are identically shaped (both are infinite half-spaces). Within this formalism we can readily compute the adhesion energies and hence the surface free energies. As the surface tensions are defined for flat interfaces between two semi-infinite phases, these energies can be expressed in terms of integrals over $\Pi$ and one finds \cite{DMAS11}:
\begin{eqnarray}
{\cal A}_{\rm AB}=-\int_0^\infty\Pi_{\rm AB}(z)\,dz&=&\gamma_A+\gamma_{B}-\gamma_{\rm AB}\label{eq:AB}\\
{\cal A}_{AA}=-\int_0^\infty \Pi_{AA}(z)\,dz&=&2\gamma_{A}\label{eq:AA}\\
{\cal A}_{BB}=-\int_0^\infty \Pi_{BB}(z)\, dz&=&2\gamma_{B}.\label{eq:BB}
\end{eqnarray}

\subsection{DFT description extended to solids}
\label{sect:boundaryconds}
We now extend the density theory description to describe soft solids such as elastomers or gels. The elastic properties in such solids are of entropic origin and their structure is close to that of a liquid. Therefore, we hypothesize that these solids can be described using the Density Functional Theory in the sharp interface approximation, by replacing the repulsive pressure $-p_r$ by an elastic stress tensor $\bar\sigma$. Just like the liquid in the previous section, this solid is assumed to be submitted to the long-range attractive potentials $\phi_{LS}$ and $\phi_{SS}$, which are not included in the stress tensor $\bar \sigma$. We therefore get a very similar expression as in the liquid case:
\begin{equation}
\label{eq:miceqsolid}
\nabla \cdot \bar\sigma - \nabla (\phi_{SS}+\phi_{SL}) =0\;,
\end{equation}
as anticipated in Eq.~\eqref{eq:miceq}.

The next step is to find expressions for the components of $\bar\Sigma$ and $\bar\sigma$, which are generally not isotropic in the elastic case. The description of the solid differs from the liquid discussed in the previous section by the existence of a base state with respect to which deformations are measured. One standardly considers, for bulk elasticity, a base state which is stress free. However, this choice is problematic for capillary effects as these induce a pre-stress, even when no external force is exerted on the solid. To follow experimental constraints, we will therefore consider that the base state is the state in which the system was prepared. If a gel is prepared in a solid mold as in~\cite{MPFPP10,MDSA12}, the base state will therefore be sensitive to the molecular interactions between this mold and the gel. For the sake of simplicity, we will reduce the discussion to the case were the solid is prepared in void. 

In order to express the constitutive relation of the material, we split both stresses into a reference part that is associated with the base state, and the  elastic contribution relative to this base state:
\begin{eqnarray}
\bar\Sigma &=& \bar\Sigma^\mathrm{ref}+\bar\Sigma^\mathrm{el} , \label{eq:splitrefandelmac}\\
\bar\sigma &=& \bar\sigma^\mathrm{ref}+\bar\sigma^\mathrm{el} 
\label{eq:splitrefandelmic}\;.
\end{eqnarray}
$\bar\Sigma^\mathrm{ref}$ and $\bar\sigma^\mathrm{ref}$ are the stresses in a reference state, taken as a semi-infinite solid in contact with vacuum (i.e. a solid-vacuum interface). For $\bar\Sigma\sel$ and $\bar\sigma\sel$ we apply Hooke's law. We will write the expressions using the thermodynamic stress $\Sigma_{ij}^\textrm{el}$ only, but the same relations hold for the microscopic stress $\sigma_{ij}^\textrm{el}$. Hooke's law relates the components of the stress $\bar\Sigma^\mathrm{el}$ to the elastic strain $\bar\varepsilon$:
\begin{equation}
\label{eq:hooke}
{\Sigma}_{ij}^\textrm{el} =  \frac{E}{1 + \nu} \left[  \varepsilon_{ij} + \frac{\nu}{1 - 2 \nu}  \varepsilon_{ll} \delta_{ij} \right]
\end{equation}
or equivalently:
\begin{equation}
 \varepsilon_{ij}=\frac{1+\nu}{E}  \Sigma_{ij}^\textrm{el}-\frac{\nu}{E} \Sigma_{kk}^\textrm{el}\delta_{ij}
\end{equation}
where
\begin{equation}
 \varepsilon_{ij} =\frac 12\;\left(\partial_i u_j+\partial_j u_i\right)\;,
\end{equation}
with $u_i$ is the displacement vector, and $E$ the Young's modulus. We consider plane strain conditions ($\varepsilon_{xx}=\varepsilon_{yy}=0$, with the $z$-direction normal to the interface), and therefore find:
\begin{eqnarray}
\Sigma_{xx}^\textrm{el}=\frac{\nu}{1-\nu}\Sigma_{zz}^\textrm{el}.
\end{eqnarray}
As explained above, the same expression holds for $\sigma_{ij}^\textrm{el}$, so:
\begin{equation}
\label{eq:hookemic}
\sigma_{xx}^\textrm{el}=\frac{\nu}{1-\nu}\sigma_{zz}^\textrm{el}.\\
\end{equation}
This relation between the $xx$ and $zz$ components of the stress-tensor is at the core of the inequality between $\gamma_{SL}$ and $\Upsilon_{SL}$. Note that the stress-tensor for an incompressible surface layer ($\nu=1/2$) is isotropic ($\sigma\sel_{xx}=\sigma\sel_{zz}$) and we would recover liquid behaviour: $\gamma_{SL}=\Upsilon_{SL}$

\subsection{Equilibrium conditions at planar interfaces}
\begin{figure}[ht!]
\includegraphics{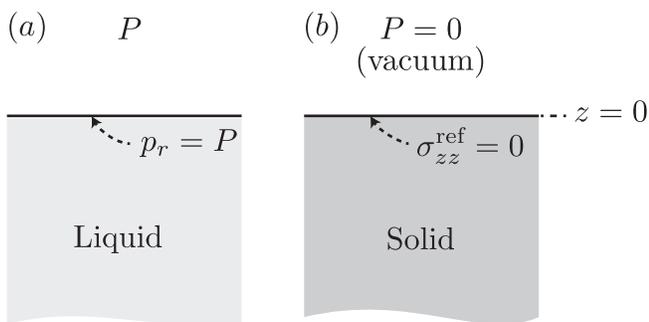}
\caption{\label{fig:eqs} 
Schematic of a liquid and solid interface. (a) A liquid-vapour interface located at $z=0$. The vapour above the liquid is at pressure $P$. Due to the low density of the vapour phase, this pressure $P$ consists entirely of short-ranged repulsive (kinetic) stress such that $p_r=P$. Since $p_r$ is continuous at the interface this can be used as a boundary condition for Eq.~\eqref{eq:liquid}. (b) A solid-vacuum interface, which serves as the reference system without elastic stresses ($\bar\Sigma\sel=\bar\sigma\sel=\vec 0$).
Similarly as in the liquid, this system fixes the value of $\sigma\sref_{zz}=-P=0$ at $z=0$, providing the boundary condition for Eq.~\eqref{eq:miceqsolid}.
}
\end{figure}
We now consider the equilibrium near a planar interface, as sketched in Fig.~\ref{fig:eqs}. First, we note that the macroscopic equilibrium condition for the fluid reads:
\begin{equation}
\label{eq:liqequil}
\Sigma_{ij} = -P\delta_{ij} \;\mbox{~everywhere,}
\end{equation}
where $P$ is the thermodynamic pressure. In the liquid case we find that, cf. Eq.~\eqref{eq:liquid}:
\begin{equation}
\label{eq:liqhom}
p_r+\phi_{LL}+\phi_{SL}=cnst\;.
\end{equation}
The unknown constant is set by both equilibrium conditions -- Eqs.~\eqref{eq:liqequil} and \eqref{eq:liqhom} -- at the liquid-vapour (or equivalently for a liquid-liquid) interface, Fig.~\ref{fig:eqs}(a), with the vapour phase at pressure $P$. At the liquid-vapour interface [$z=0$, cf. Fig~\ref{fig:eqs}(a)] the local potential due to liquid-liquid interactions reads $\phi_{LL}=\Pi_{LL}(0)$ and due to the absence of any solid $\phi_{SL}=0$. The vapour phase above the liquid is of such low density that only the repulsive contribution applies, and therefore we find in the vapour $p_r=P$.  Furthermore, microscopic the stress tensor must be continuous at $z=0$, which implies $p_r=P$ at $z=0$. This allows us to compute the integration constant in Eq.~\eqref{eq:liqhom}, and we obtain
\begin{equation}
\label{eq:prliq}
p_r+\phi_{LL}+\phi_{SL}=P+\Pi_{LL}(0)\;,
\end{equation}
everywhere within the liquid.

\begin{figure*}[t!]
\includegraphics{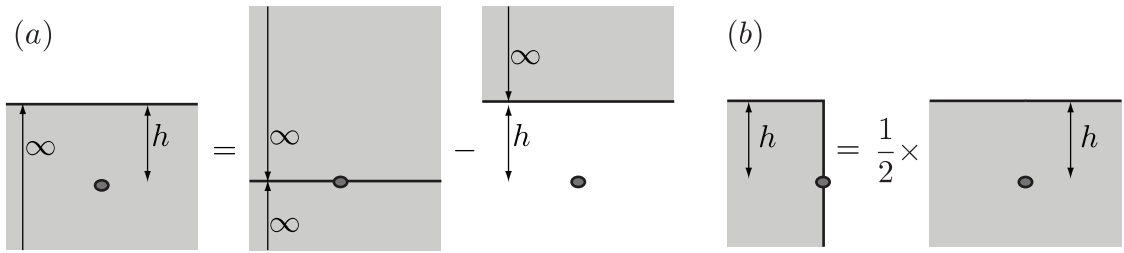}
\caption{\label{fig:intdomains} Schematics on the construction of the integration domains required to evaluate the potential on the edge of the control volume (Fig.~\ref{fig:sketch}) due to the absence of a semi-halfspace of material $\alpha$ at distance $h$. (a) The potential at the indicated point in the condensed phase (at distance $h$ from the vacuum interface) is constructed by subtracting the potential due to a semi-infinite volume of the condensed phase at distance $h$ [$\Pi_{\alpha\alpha}(h)$] from the potential of a complete, infinite volume of the condensed phase [$2\Pi_{\alpha\alpha}(0)$]. (b) To evaluate the potential at the edge of the control volume as indicated in Fig.~\ref{fig:sketch}, we only take into account the potential due to material {\em outside} this control volume, which is a quarter-space, thus only half of the result from (a). Finally, one obtains $\Pi_{\alpha\alpha}(0)-\Pi_{\alpha\alpha}(h)/2$.}
\end{figure*}

We now have an explicit relation between the thermodynamic pressure $P$ (or stress $\bar\Sigma$) and the microscopic contributions to the pressure $p_r$ and $\phi$. 
In planar geometries, we recover that in the liquid phase $\phi_{LL}=2\Pi_{LL}(0)-\Pi_{LL}(z)$ [Fig.~\ref{fig:intdomains}(a)] where the $2\Pi_{LL}(0)$-term represents the attractive interactions due to the surrounding liquid in the bulk, and the $\Pi_{LL}(z)$-term the missing interactions (compared to the bulk) due to the presence of an interface at distance $z$. We then find, for the liquid, using Eq.~\eqref{eq:prliq}:
\begin{equation}
\label{eq:liqpr2}
p_r=P-\Pi_{LL}(0)+\Pi_{LL}(z)-\phi_{SL}\;,
\end{equation}
In the bulk of the liquid, where $\Pi_{LL}(z\to\infty)=0$ and $\phi_{SL}=0$, $p_r$ tends to:
\begin{equation}
\label{eq:liqprbulk}
p_r^{bulk}=P-\Pi_{LL}(0)\;.
\end{equation}

%%ELASTIC SOLID
Finally, we derive the microscopic equilibrium equation for the elastic solid. We define the base state as the state where the solid is in contact with vacuum. By definition, the elastic stress $\bar\sigma\sref$ then vanishes. We follow the same path as in the liquid case, with the distinction that in the solid case the stress tensor is not isotropic. At the solid-vacuum interface [$z=0$, cf. Fig.~\ref{fig:eqs}(b)] we find $\sigma_{zz}\sref = 0$, because a vacuum represents $P=0$. We use this boundary condition to determine the integration constant in Eq.~\eqref{eq:miceqsolid}, and find that:
\begin{equation}
\label{eq:sigmazzref}
\sigma_{zz}^\textrm{ref} = \Pi_{SS}(0) - \Pi_{SS}(z) ,
\end{equation}
where we used that $\phi_{SS}$ anywhere in the solid phase at distance $z$ from the interface is given by $2\Pi_{SS}(0)-\Pi_{SS}(z)$, cf. Fig.~\ref{fig:intdomains}(a). Furthermore, since $\nabla\cdot\bar\Sigma=0$ and $\Sigma_{zz}(z=0)=0$ we find:
\begin{equation}
\Sigma\sref_{zz}=0\;.
\end{equation}
In order to calculate the excess stress in sect.~\ref{sect:forces}, a relation between $\Sigma\sref_{xx}$ and $\sigma\sref_{xx}$ is required. Posing that the excess stress in the bulk is zero (which is the definition of an excess quantity), we find that in the bulk:
\begin{equation}
\label{eq:solrefxxbulk}
\sigma_{xx}^\textrm{ref,bulk}=\Sigma\sref_{xx}+\Pi_{SS}(0)\;.
\end{equation}
This is the solid analogue of \eqref{eq:liqprbulk} for the liquid. Although the bulk value of $\sigma_{xx}\sref$ is now known, we did not yet account for the presence of the vacuum, which is felt at small distances from an interface.  Analogously to the liquid in Eq.~\eqref{eq:liqpr2} this will give:
\begin{equation}
\label{eq:solrefxx}
\sigma_{xx}\sref=\Sigma\sref_{xx}+\Pi_{SS}(0)-\Pi_{SS}(z)\;.
\end{equation}
Note that  the term $-\Pi_{SS}(z)$ does not change the value of the stress in the bulk, Eq.~\eqref{eq:solrefxxbulk}, as it vanishes at large $z$. We also anticipate that we do not require an explicit expression for $\Sigma\sref_{xx}$, as it will cancel out when computing the excess stress as described in the next section.

\section{Excess quantities and surface stress}
\label{sect:forces}
In this section, we derive the surface stress by calculating the total excess stress across the interface. Introducing the coordinate $z$ normal to the interface the interfacial excess quantity of an extensive measure $M$ is defined as
\begin{equation}
{\cal M}_{\rm AB} =S_{\rm AB}\int \left[m(z)-m^A \theta(z)-m^B\theta(-z)\right] dz\;,
\end{equation}
with $m(z)$ the corresponding intensive quantity and $\theta$ the Heaviside step-function. The total excess ${\cal M}$ is therefore the (integrated) difference between the bulk values of $m^A$ and $m^B$ and the true value of $m(z)$ that varies continuously in the vicinity of the interface. Throughout the analysis we will use superscripts to denote the bulk value of a given phase. Thermodynamic quantities such as mass density, entropy or energy exhibit an interfacial excess. For example, the surface free energy $\gamma_{\rm AB}$ can be evaluated from the microscopic interactions using the classical relations (\ref{eq:AB}-\ref{eq:BB}).

\begin{figure}[ht!]
\includegraphics{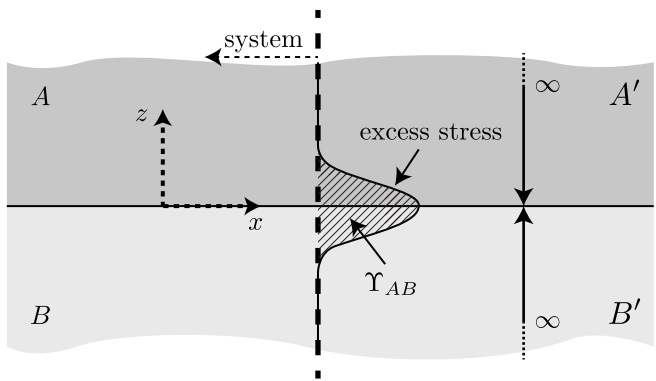}
\caption{\label{fig:sketch} Schematic of a planar interface between two semi-infinite phases $A$ and $B$. Near the interface, an excess stress develops and the total (integrated) excess stress is defined as the surface stress $\Upsilon_{\rm AB}$. Note that, by definition, the excess stress is zero in the bulk (far away from the interface). To evaluate the surface stress associated with this interface $\Upsilon_{\rm AB}$ we define a control surface perpendicular to the $AB$-interface, across which the excess stress on the control volume (left) due to contact forces and long-range forces originating from outside the control volume is integrated.}
\end{figure}

In this section we calculate the surface stress $\Upsilon_{\rm AB}$, which is depicted schematically in Fig.~\ref{fig:sketch}. The surface stress is defined as the integrated excess stress, and has the dimension of a force per unit length, or equivalently of an energy per unit area. While for liquid-vapor or liquid-liquid interfaces the surface stress $\Upsilon_{\rm AB}$ is identical to the surface free energy $\gamma_{\rm AB}$ (following from the virtual work principle), this is not the case when one of the phases is elastic. In this section we provide explicit expressions for $\Upsilon_{\rm AB}$, starting from the definition

\begin{equation}\label{eq:def}
\mathcal{F}_x = \Upsilon_{\rm AB} + \int_0^\infty dz \Sigma_{xx}^A + \int_{-\infty}^0 dz \Sigma_{xx}^B,
\end{equation}
where $\Sigma_{ij}^A$ and $\Sigma_{ij}^B$ are thermodynamic bulk stresses; note that these thermodynamic stresses implicitly contain all microscopic interactions (short- and long-ranged), so that for the liquid one simply recovers the thermodynamic pressure $\Sigma_{ij}=-P\delta_{ij}$. $\mathcal{F}_x$ is the total force per unit length acting on the control volume $AB$, left of the dashed line in Fig.~\ref{fig:sketch}, projected parallel to the interface. This force is exerted by the volume indicated $A'B'$, right of the dashed line in Fig.~\ref{fig:sketch}. By computing this force $\mathcal{F}_x$ explicitly from the microscopic models described in the previous section, we will obtain explicit expressions for $\Upsilon_{\rm AB}$ for all combinations of liquid/vapour, liquid/liquid, and solid/liquid interfaces. 

\subsection{Force calculation}
The interactions in the DFT models consist of a short-ranged contact stress $\sigma_{ij}$, and long-ranged potential interactions $\phi_{\alpha \beta}$. The total mechanical force on the control volume therefore reads
\begin{eqnarray}
\mathcal{F}_x &=&  \int_0^\infty dz \left(\sigma_{xx} - \phi_{\rm A'A} - \phi_{\rm B'A}\right) \nonumber \\
&&+\int_{-\infty}^0 dz \left(\sigma_{xx} - \phi_{\rm B'B} - \phi_{\rm A'B}\right),
\end{eqnarray}
where we integrated the the long-ranged forces per unit volume, $-\partial \phi_{\alpha\beta}/\partial x$, up to the dashed line in Fig.~\ref{fig:sketch}. The terms $\phi_{\rm A'A}$ represents the potential induced by volume $A'$ (outside the control volume) on the part of the same phase that lies within the control volume. Note the subtle difference with $\phi_{AA}$ appearing in the equilibrium for $p_r$ which is the potential energy due to the entire volume. Combined with the definition (\ref{eq:def}), we find the surface stress:
\begin{eqnarray}
\Upsilon_{\rm AB} &=&  \int_0^\infty dz \left(\sigma_{xx} - \phi_{\rm A'A} - \Sigma_{xx}^A\right) \nonumber \\
&&+\int_{-\infty}^0 dz \left(\sigma_{xx} - \phi_{\rm B'B} - \Sigma_{xx}^B\right) \nonumber \\
&& -\int_0^\infty dz  \,\phi_{\rm B'A} - \int_{-\infty}^0 dz \,\phi_{\rm A'B}.\label{eq:poep}
\end{eqnarray}
As the long-ranged potentials $\phi_{\alpha\beta}$ follow from volume integrals [see Eq.~(\ref{eq:phi})], their contribution is completely independent of the rheology of the materials and can be readily evaluated. In addition, all the energies  can be expressed directly in terms of $\Pi_{\alpha\beta}$, for which phase $\alpha$ is a flat semi-infinite space, owing to the symmetry of the domains in Fig.~\ref{fig:sketch}. Let us first exploit this connection for the ``cross terms'' $B'A$ and $A'B$. From the geometry one can see that $\phi_{\rm B'A}=\frac{1}{2} \Pi_{\rm BA}$, due to a missing quadrant of phase $B$. Since by symmetry $\phi_{\rm A'B}=\frac{1}{2} \Pi_{\rm AB}=\frac{1}{2} \Pi_{\rm BA} $, the last two terms in (\ref{eq:poep}) combine to an integral over $\Pi_{\rm AB}$ and according to (\ref{eq:AB}) this yields the adhesion energy ${\cal A}_{\rm AB} = \gamma_A +\gamma_B -\gamma_{\rm AB}$. The terms $A'A$ and $B'B$ are slightly more difficult to interpret, as can be seen from Fig.~\ref{fig:intdomains}. As explained in detail in the caption of the figure, one finds $\phi_{\rm A'A}=\frac{1}{2}(2\Pi_{AA}(0)-\Pi_{AA}(z))$. 

Using the above expressions for the potentials, we can split the surface stress \eqref{eq:poep} into three contributions:
\begin{eqnarray}
\Upsilon_{\rm AB} &=&  \Upsilon_{\rm AB}^A + \Upsilon_{\rm AB}^B + \left( \gamma_A +\gamma_B -\gamma_{\rm AB}\right),
\label{eq:poep2}
\end{eqnarray}
where 
\begin{subequations}
\label{eq:poep3}
\begin{align}
\Upsilon_{\rm AB}^A &=&  \int_0^\infty dz \left( \sigma_{xx} - \Pi_{AA}(0)+\frac{1}{2}\Pi_{AA}(z) - \Sigma_{xx}^A \right), \label{eq:poep3a}\\
\Upsilon_{\rm AB}^B &=&  \int_{-\infty}^0 dz \left( \sigma_{xx} - \Pi_{BB}(0)+\frac{1}{2}\Pi_{BB}(-z) - \Sigma_{xx}^B \right)\nonumber \\
&=&\int_{0}^\infty dz \left( \sigma_{xx} - \Pi_{BB}(0)+\frac{1}{2}\Pi_{BB}(z) - \Sigma_{xx}^B \right).\label{eq:poep3b}
\end{align}
\end{subequations}
The last contribution of (\ref{eq:poep2}) is the work of adhesion; it is completely independent of the elastic properties of phases $A$ and $B$. By contrast, the direct interactions $\Upsilon_{\rm AB}^A$ and $\Upsilon_{\rm AB}^B$ involve the stress $\sigma_{xx}$, which has to be determined from the constitutive relations.

\subsection{The liquid/vapor interface: $\Upsilon_{LV}=\gamma_{LV}$}
From Shuttleworth's thermodynamic relation, it follows that for incompressible liquids there is a strict equality between surface energy and surface stress \cite{Shuttleworth50,WAS13}. Hence, one usually does not distinguish between the two concepts, and simply use the nomenclature of surface tension. As a consistency check for the formalism that is developed in this paper, we first verify explicitly that this equality is recovered for a liquid/vapor interface, i.e. $\Upsilon_{LV}=\gamma_{LV}$ and a liquid/liquid interface, i.e. $\Upsilon_{\rm AB}=\gamma_{\rm AB}$ with $A$, $B$ liquids.
Then, we will study the particular case of a liquid in contact with an elastic solid, for which we will see that in general $\Upsilon_{SL}\ne\gamma_{SL}$.

Let us therefore consider phase $A$ to be vacuum and phase $B$ to be a liquid. The thermodynamic stress inside the liquid is isotropic, i.e. $\Sigma_{xx}^B=-P$. For a true vacuum the thermodynamic pressure is strictly zero, but in real physical systems the liquid will equilibrate at a small nonzero vapor pressure. For such a liquid/vapor interface, the surface energies become $\gamma_{A}=0$, $\gamma_{B}=\gamma_{LV}$ and $\gamma_{\rm AB}=\gamma_{LV}$. As a consequence of the vacuum, the work of adhesion $\gamma_{A}+\gamma_{B}-\gamma_{\rm AB}=0$, and also $\Upsilon_{\rm AB}^A=0$. Hence, the total surface stress $\Upsilon_{LV}$ only receives a contribution from liquid-liquid interactions,  $\Upsilon_{LV} = \Upsilon_{LV}^L$, as defined in (\ref{eq:poep3}).

The integral (\ref{eq:poep3b}) involves $\sigma_{xx}$, which from Eqs.~(\ref{eq:liqhom}, \ref{eq:liqpr2}) can be written

\begin{eqnarray}
\sigma_{xx} &=& - p_r = -P - \Pi_{LL}(0) + \phi_{LL} \nonumber \\
&=& -P + \Pi_{LL}(0) - \Pi_{LL}(-z).
\end{eqnarray}
Inserting this in (\ref{eq:poep3b}), one finds a simplified expression 

\begin{equation}
\Upsilon_{LV}^V = -\frac{1}{2} \int_{-\infty}^0 dz \Pi_{LL}(-z) = -\frac{1}{2} \int_0^\infty dz \Pi_{LL}(z). 
\end{equation}
According to (\ref{eq:AA}), the last integral can indeed be identified with the surface energy $\gamma_{LV}$. This completes the demonstration that the mechanical excess stress at the liquid/vapor interface, defined in Fig.~\ref{fig:sketch} and computed from the DFT model, is strictly equal to the surface energy: $\Upsilon_{LV}=\gamma_{LV}$. 

\subsection{The liquid/liquid interface: $\Upsilon_{\rm AB}=\gamma_{\rm AB}$}

We now consider an interface between two immiscible liquids $A$ and $B$, so again the stress tensor will be isotropic $\Sigma_{xx}^A=\Sigma_{xx}^B=-P$. Retracing the steps of the analysis for the liquid/vapor interface, the main difference is that the equilibrium equation for the repulsive pressure $p_r$ now has a contribution $\phi_{\rm AB}=\Pi_{\rm AB}$. Hence, one finds in phase $A$:

\begin{eqnarray}
\sigma_{xx} &=& -p_r = -P + \Pi_{AA}(0) -\Pi_{AA}(z) + \Pi_{\rm AB}(z),\nonumber\\
\end{eqnarray}
while in phase $B$
\begin{eqnarray}
\sigma_{xx} &=& -p_r = -P + \Pi_{BB}(0) - \Pi_{BB}(z) + \Pi_{\rm AB}(-z).\nonumber\\
\end{eqnarray}
Note that this result is equivalent to Eq.~\eqref{eq:liqpr2}, with the solid phase replaced by the other liquid.
This gives for the integrals (\ref{eq:poep3})

\begin{align*}
\Upsilon_{\rm AB}^A=\gamma_A-(\gamma_A+\gamma_B-\gamma_{\rm AB})=\gamma_{\rm AB}-\gamma_{B},\\
\Upsilon_{\rm AB}^B=\gamma_B-(\gamma_A+\gamma_B-\gamma_{\rm AB})=\gamma_{\rm AB}-\gamma_{A} .
\end{align*}
Adding the work of adhesion, one indeed obtains that the surface stress is equal to the surface energy:

\begin{equation}
\Upsilon_{\rm AB} = \Upsilon_{\rm AB}^A+\Upsilon_{\rm AB}^B + \gamma_A +\gamma_B - \gamma_{\rm AB} = \gamma_{\rm AB}.
\end{equation}

\subsection{The solid/liquid interface}
Finally, we discuss the solid-liquid interface.
As before, we evaluate the terms in the integral \eqref{eq:poep3} seperately.
The elastic properties of the solid enter through $\Upsilon^S_{LS}$, whereas $\Upsilon^L_{LS}$ is the same as that in the liquid/liquid case,
\begin{equation}
\Upsilon^L_{LS}=\gamma_{SL}-\gamma_{SV}\;.
\end{equation}
To obtain $\sigma_{xx}$ in the equation for $\Upsilon^S_{LS}$ [Eq.~\eqref{eq:poep3b}] we use once more that the normal stress $\sigma_{zz}$ is continuous across the liquid-solid inteface. 
At the liquid side we know that $\sigma_{zz}(z=0)=-p_r=-P+\Pi_{SL}(0)$, which allows us to determine the integration constant in Eq.~\eqref{eq:miceqsolid} and we obtain, for the solid:
\begin{eqnarray}
\sigma_{zz}(z)&=&-P+\Pi_{SS}(0)+\Pi_{SL}(z)-\Pi_{SS}(z)\;,
\end{eqnarray}
To obtain the elastic response due to the presence of the liquid, we subtract the reference stress (solid in contact with vacuum, Eq.~\eqref{eq:sigmazzref}) and obtain:
\begin{equation}
\sigma_{zz}\sel(z)=\Pi_{SL}(z)-P\;.
\end{equation}
Compared to the base state (Sect.~\ref{sect:boundaryconds}), the liquid interactions $\Pi_{SL}(z)$ are added to the pressure $P$.
Then, via Hooke's law under plane strain conditions in the $x$- and $y$-directions we get (Eq.~\eqref{eq:hookemic}):
\begin{equation}
\sigma_{xx}\sel(z)=\frac{\nu}{1-\nu}(\Pi_{SL}(z)-P)
\end{equation}
and thus, following Eqs.~\eqref{eq:splitrefandelmic} and \eqref{eq:solrefxx}:
\begin{equation}
\sigma_{xx}(z)=\Sigma\sref_{xx}+\Pi_{SS}(0)-\Pi_{SS}(z)+\frac{\nu}{1-\nu}(\Pi_{SL}(z)-P).
\end{equation}
The final step before we can evaluate $\Upsilon^S_{SL}$ is to determine $\Sigma^S_{xx}$ by applying Hooke's law on $\Sigma\sel_{zz}=-P$:
\begin{equation}
\Sigma^S_{xx}=\Sigma_{xx}\sref-\frac{\nu}{1-\nu}P\;.
\end{equation}
We can now evaluate $\Upsilon^S_{SL}$, Eq.~\eqref{eq:poep3b}, and find:
\begin{eqnarray}
\Upsilon_{LS}^S&=&\int_{-\infty}^0 \left( -\frac{\Pi_{SS}(z)}{2}+\frac{\nu}{1-\nu}\Pi_{SL}(z)\right) dz\nonumber\\
&=&\gamma_{SV}-\frac{\nu}{1-\nu}(\gamma_{SV}+\gamma_{LV}-\gamma_{SL})\;,
\end{eqnarray}
and we obtain for the total surface stress:
\begin{equation}
\nonumber\Upsilon_{SL}=\frac{\nu}{1-\nu}\gamma_{SL}+\frac{1-2\nu}{1-\nu}(\gamma_{SV}+\gamma_{LV})
\end{equation}
which is the central result of this paper, Eq.~\eqref{eq:upsilon_SL}.

\section{Discussion}
\label{sect:disc}
\begin{figure}[ht!]
\includegraphics{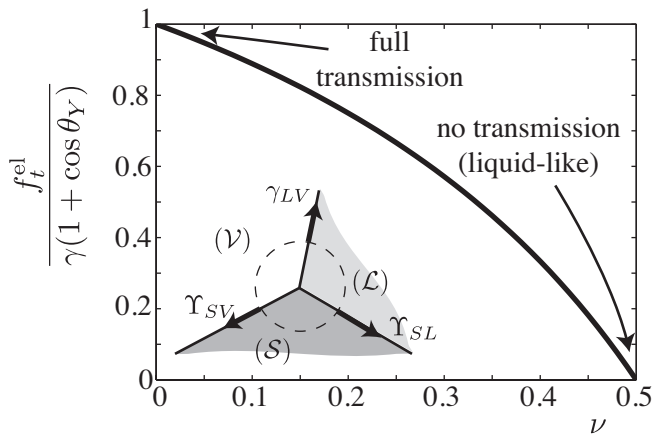}
\caption{\label{fig:nurftel} Dependence of the transmitted tangential stress on the superficial Poisson ratio $\nu$: Eq.~\eqref{eq:pies}. 
The dashed circle indicate a control volume around the contact line, which is larger than the molecular interaction range.
For $\nu=0$, the full stress on the solid induced by the presence of the contact line is transmitted to the solid bulk, where it is balanced by elastic effects. In the incompressible limit $\nu=1/2$, no tangential stress is transmitted. This must necessarily hold for liquids, which can offer no elastic resistance to any transmitted tangential stress.}
\end{figure}
The Shuttleworth equation \eqref{POEPIE} is a thermodynamic relation showing that one needs to distinguish between the surface stress $\Upsilon$, i.e. the excess force per unit length in the interface, and the surface free energy $\gamma$. By computing the elastic and enthalpic interactions from a microscopic model, we have for the first time derived an explicit relation between $\Upsilon$ and $\gamma$. We find that, in general, the surface stress of a solid interface is indeed different from the surface energy of the same interface, as is quantified by Eq.~\eqref{eq:upsilon_SL}. It turns out that the difference between surface stress and surface energy depends on the compressibility in the superficial zones of the solid-liquid interface, characterized by the Poisson's ratio $\nu$. The relation shows that for solids that are incompressible in the superficial zone ($\nu=1/2$) the surface stress is equal to the surface energy, in line with the fact that for incompressible liquids $\gamma=\Upsilon$. The model predicts an equality $\gamma_{SV}=\Upsilon_{SV}$ for a solid-vapor interface of arbitrary value of $\nu$ [consider \eqref{eq:upsilon_SL} with $L=V$ and $\gamma_{VV}=0$]. 

In our previous work \cite{WAS13} we have shown that the difference between surface stress and surface energy is crucial for understanding the elastic deformation of partially wetted deformable surfaces. In particular, it was demonstrated that the contact line region transmits a residual tangential force to the substrate, which has to be balanced by bulk elasticity. The magnitude of this tangential force can be expressed as (see inset Fig.~\ref{fig:nurftel}):
\begin{equation}
f_t^{el} = \left(\Upsilon_{SL} - \Upsilon_{SV} \right) - \left( \gamma_{SL} - \gamma_{SV}\right).
\end{equation}
With \eqref{eq:upsilon_SL} this becomes:
\begin{equation}
\label{eq:pies}
f_t^{el}=\frac{1-2\nu}{1-\nu}\gamma(1+\cos\theta)\;.
\end{equation}

The key feature of this result is that a finite tangential force is transmitted whenever the Poisson ratio $\nu \neq 1/2$, see Fig.~\ref{fig:nurftel}. 
We can directly compare this result with recent force measurements performed in molecular dynamics simulations~\cite{WAS13}.
For such a cubic lattice with springs of equal spring constant between nearest and next-nearest neighbors, it can be shown that $\nu=0.2$.
In these simulations, parameters were chosen such that $\gamma_{SL}=\gamma_{SV}$, so $\cos\theta=0$. 
The numerically measured value of the transmitted tangential stress ($f^\textrm{el}_t/\gamma=0.81\pm 0.17$) agrees quantitatively with Eq.~\eqref{eq:pies}, which predicts $f^\textrm{el}_t/\gamma=0.75$ for such a system.

The influence of the tangential force recovered in Eq.~\ref{eq:pies} is usually not taken into account when describing elasto-capillary deformations \cite{BicoNATURE,PRDBRB07,BoudPRE,PyEPJST,HonsAPL,RomanJPCM,ChiodiEPL,HurePRL,L61,Rusanov75,Yuk86,Shanahan87,CGS96,White03,PericetCamara08,PBBB08,MPFPP10,SARLBB10,LM11,JXWD11,,SBCWWD13,WBZ09,Style12,Limat12}. One might argue that for most experiments this is a valid assumption: soft rubbers and gels are essentially incompressible, in which case our model recovers $f_t^{el}=0$. Intriguingly, however, recent experiments on a soft elastomeric wire clearly demonstrate that a tangential force is transmitted across the surface, and give rise to a significant tangential bulk strain \cite{MDSA12}. This strongly suggests that, while the bulk material clearly has $\nu \approx1/2$, the interfacial layers of soft elastomers \emph{do} exhibit a substantial degree of compressibility. This would explain why a tangential stress is transmitted in the vicinity of the contact line, even for incompressible bulk materials. A final consequence of such a tangential force is that the contact angles on very soft surfaces do not follow Neumann's law \cite{MDSA12b}. This provides a strong limitation on the recently proposed method to determine surface stresses from contact angle measurements \cite{Style13}. Our work clearly shows that a more detailed microscopic description of the interfacial chemistry, combined with further experimental characterizations, are necessary to fully resolve the physics of elastocapillarity. 

The authors acknowledge input by and discussions with Antonin Marchand.
%\bibliographystyle{prsty_withtitle}
%\bibliography{all_ref}

\end{document}